# A Qualitative Representation and Similarity Measurement Method in Geographic Information Retrieval

Yong Gao[1], Lei Liu[1], Xing Lin[1] Yu Liu[1] *

[1] Institute of Remote Sensing and Geographic Information Systems, Peking University, Beijing 100871, China

*Corresponding author, e-mail: liuyu@urban.pku.edu.cn

**Abstract** The modern geographic information retrieval technology is based on quantitative models and methods. The semantic information in web documents and queries cannot be effectively represented, leading to information lost or misunderstanding so that the results are either unreliable or inconsistent. A new qualitative approach is thus proposed for supporting geographic information retrieval based on qualitative representation, semantic matching, and qualitative reasoning. A qualitative representation model and the corresponding similarity measurement method are defined. Information in documents and user queries are represented using propositional logic, which considers the thematic and geographic semantics synthetically. Thematic information is represented as thematic propositions on the base of domain ontology. Similarly, spatial information is represented as geo-spatial propositions with the support of geographic knowledge base. Then the similarity is divided into thematic similarity and spatial similarity. The former is calculated by the weighted distance of proposition keywords in the domain ontology, and the latter similarity is further divided into conceptual similarity and spatial similarity. Represented by propositions and information units, the similarity measurement can take evidence theory and fuzzy logic to combine all sub similarities to get the final similarity between documents and queries. This novel retrieval method is mainly used to retrieve the qualitative geographic information to support the semantic matching and results ranking. It does not deal with geometric computation and is consistent with human commonsense cognition, and thus can improve the efficiency of geographic information retrieval technology.

## 1.Introduction

Geographic Information Retrieval (GIR) is an applied information search area that concerns indexing, searching, retrieving, and browsing georeferenced information sources[1]. It can improve the effectiveness of retrieving information from web documents by taking spatial location and semantics into consideration. As well as the traditional information retrieval technologies, GIR mainly handles web documents written in natural language, in which the representation model of documents and queries, and the similarity measurement between them are the basis. Different from traditional information retrieval, a GIR system emphasizes on spatial location or extent in addition to thematic information.

Current GIR systems extract, store and match thematic information and geographic information separately. The thematic information is handled by traditional information retrieval methods that take the keywords' co-occurrences between queries and documents as the thematic similarity [2]. Meanwhile, the geographic information in documents, mainly presented as place names or addresses, is transformed into geometries with coordinates under the support of gazetteers or geographic thesauri. These geometries are taken as the document geographic footprint that can be represented as many different forms, such as centroid, minimum bounding rectangle, convex hull, and polygon [3-6]. Similarly, spatial restrictions in queries are also converted into exact geographic footprints. Hence, the spatial similarity is obtained by calculating the overlapping areas or distances between the footprints of queries and documents [7]. At last, the final similarities are got by combining thematic similarities and spatial similarities respectively [8, 9] to rank the candidates.

The present representation and similarity measurement methods simplify the geographic information processing, but there are some problems during the process. They handle thematic information and geographic information separately, ignoring the semantic relationship between them, and are therefore not adaptive to human spatial cognition and may make candidate documents unmatch what user needs. In fact, the thematic and geographic information of an event should be treated as a whole. On the other hand, the similarity metrics, including Euclidean distance, and overlap areas of geometries, ignore semantics between places and may lead mismatch results.

To avoid the drawbacks of conventional GIR, a qualitative representation and similarity measurement method is proposed in this paper based on proposition logic for both thematic and geographic information. The reasoning procedure is more consistent with to human spatial recognition, so it can understand geographic queries and documents better and retrieve information needed more efficiently.

## 2.Related Work

Geographic information retrieval is the extension of traditional information retrieval that accesses information that relates to specific geographic locations [10]. It includes all of the areas of traditional information retrieval, but in addition has an emphasis of spatially-oriented indexing and retrieval [1]. Previous research in GIR has addressed problems such as the representation model of geographic information, retrieval model and similarity metrics. Many representation models for geographic information are available and they are relevant to the sequent similarity metrics.

Anastacio et al. [11] compared different approaches for assigning geographic scopes to documents and found that overall the Web-a-Where method gives the best experiment values. This method is closely followed by a simple baseline that assigns the most frequent place reference as the geographic scope. Martins et al. [12] tested two different approaches at GeoCLEF 2006, namely the relatively simple augmentation of geographic terms in the topics, through the use of a geographic ontology, and a text mining approach based on extracting geographic references from documents, in order to assign each to a corresponding geographic scope. Frontiera et al. [5] compared several geographic similarity methods on two different representation models: minimum bounding box and convex hull. They found that although the convex hull representation may bring higher quality results but it is difficult to implement and in a large document collections it is hard to index and maintenance these documents. For most situations, the accuracy of search results of the simple minimum bounding box model is qualified enough. Palacio et al. [13] defined an evaluation framework for evaluating GIR systems according to spatial, temporal and topical information. They used a test collection for validating different combining approaches with their

PIV GIR system [14] and got a 73.9% improvement over a state-of-the-art topical baseline. The results showed that the three dimensions are not redundant, but they complement each other. Janowicz et al. [15] introduced a generic framework for semantic similarity measurement that clearly separates the process of measuring similarity and finding alignable descriptors from the concrete functions used to compute similarity values for selected tuples of these descriptors. They also discussed the role of these measures in semantics-based geographic information retrieval, introduced paradigms, and showed their implementations and limitations for real user interfaces.

The abovementioned studies divide the thematic and spatial information and represent them separately so that the representation model is simplified. In the real world, however, the thematic and spatial information are as a whole and cannot be separated. And the quantitative representation model and the similarity metrics based on overlapping area or Euclidean distance ignore the semantics behind the query. For instance, document A describes "traffic accident in Miyun county, Beijing" and document B describes "traffic accident in Jixian county, Hebei". For query "traffic accident near Pingu, Beijing", in conventional GIR systems B gets a higher ranking score since the Euclidean distance of query and B is closer. However, the place in the query is in the same province of A, so A should get a higher ranking score in consideration of semantics.

To solve the separation of thematic and spatial information and the semantics loss of geographic information, we proposed a qualitative representation model based on the propositional logic and takes the two as a whole. The evidence theory and fuzzy logic is introduced as the similarity metrics for our model. Details will be discussed in the later sections.

## 3. Qualitative GIR Representation

### 3.1 Model

GIR mainly handles documents and queries. Documents such as web pages are texts written in natural language. A document includes both thematic information and geographic information. Thematic information is composed of a set of keywords, and geographic information is represented as place names, addresses, IP addresses, phone district codes, and predicates of spatial relationships or their combination etc. A GIR query can be characterized as a triplet of <*theme*, *spatial relationship*, *location*>, denoting a topic of interest in combination with a place name qualified by a spatial preposition such as near, in, or north of [16]. In GIR systems, documents and queries are modeled similarly, both consist of thematic information and geographic information.

To solve the above problems of conventional GIR, we propose a method based on propositional logic to represent both thematic information and geographic information in a unified form. A propositional logic is a formal system, in which atomic propositions can forms formulae with the connection of logic operators, and a system of inference rules and axioms allows certain formulae to be derived. A proposition is a declarative sentence that has the quality or property of being either true or false. An atomic proposition is a proposition that cannot be broken down into other simpler sentences.

For the documents to be retrieved, the texts are processed into many independent information units by text segmentation. An information unit is composed of <*thematic information, spatial information*> and represents an independent content of the text. From the point of propositional logic, an information unit is an atomic proposition and describes a fact in the real world, such as "earthquake in Wenchuan, Sichuan". It is quite different from the conventional geographic information retrieval that takes keywords and spatial scopes to represent the document information and leads to the separation of semantics between thematic and spatial information. A document can thus be represented by a set of information units, which can easily be converted into propositions based on propositional logic. Thematic information is represented by keyword sets where each keyword is a thematic proposition. Spatial information is represented by geo-spatial proposition sets where each geo-spatial proposition is composed of place names, spatial predicates and logic operators. Then the two kinds of information can be combined together by propositional logic and represented in a unified form. Similarly, queries take the same representation model.

Generally, given a document $d$, it is segmented into many independent information units according to relationships of thematic information and geographic information. Thus the document $d$ can be represented as follows:

$$d=\{u_1, u_2, u_3, \ldots, u_n\} \quad (1)$$

where $n=|d|$ is the number of information units in $d$. When $n=0$, it shows that $d$ is a blank document and has no valid information. An unit $u_i$ has two components: thematic information $t_i$ and spatial information $g_i$, denoted as:

$$u_i=(t_i, g_i), \quad i=1, 2, 3, \ldots, n \quad (2)$$

A query can also be formalized as the same form since in natural language a spatial query is usually presented as the group <*thematic information, spatial information*>. The documents and queries' representation models are uniformed into one same type.

**3.2 Qualitative Representation of Thematic Information**

Thematic information $t_i$ in an information unit $u_j$ is represented as a set of thematic propositions:

$$t_i = \{tp_i(k) | k \in N \text{ and } 1 \leq k \leq NK_i\} \quad (3)$$

where $NK_i$ is the number of keywords conveying thematic information, and $tp_i(k)$ is a thematic proposition, which can be a single keyword or a composite proposition.

On the basis of propositional logic, a thematic proposition sentence is defined as the composition of propositions connected with logic operators $\{\wedge, \vee, \neg\}$, and its inference rules are the following:

(1) A single keyword is an atomic proposition;
(2) If $P$ is a thematic proposition sentence, its negation $\neg P$ is also a thematic proposition sentence;
(3) If both $P$ and $Q$ are thematic proposition sentences, their conjunction $P \wedge Q$ is also a thematic proposition sentence;
(4) If both $P$ and $Q$ are thematic proposition sentences, their disjunction $P \vee Q$ is also a thematic proposition sentence;

Generally, an information unit may have multi keywords: $t_i=\{k_1, k_2, k_3, \ldots, k_n\}$. For documents, thematic propositions are always built up by conjunctions of keywords, so only the first rule applies. The other three rules are set for queries, since the number of keywords is quite less and their combinations are based on Boolean logic.

In this model, domain ontologies are needed to represent and formulate thematic information in documents. It can help to extract thematic keywords and calculate semantic similarity between them. After word segmentation of documents, terms in the domain ontology are employed to recognize and match the keywords. Then conceptions in the ontology are used to formalize the representation of information in the documents.

**3.3 Qualitative Representation of Geographic information**

The geographic information $g_i$ in an information unit $u_j$ is characterized as a proposition describing geographic location, which is called geo-spatial proposition and represented as follows.
$$g_i = \{gp_i(m) | m \in N \text{ and } 1 \leq m \leq NG_i\}$$
A geo-spatial proposition $gp_i(m)$ is an expression describing geographic locations or extents, usually a phrase containing place names, such as "north of Beijing", and $NG_i$ is the number of toponyms in the information unit $u_i$.

A geo-spatial proposition can be decomposed into many atomic geo-spatial propositions [17], each of which consists of a geo-spatial predicate standing for a spatial relationship and a place name standing for a landmark. If the geo-spatial predicate is the term "equal to", it can be omitted. Geo-spatial propositions are built up by conjunctions or nests of logic operators to represent the spatial scope qualitatively.

The elements of a geo-spatial proposition sentence include propositions, logic operators $\{\wedge, \vee, \neg\}$, and spatial relationship operators $\oint$. The spatial relationship operator $\oint$ expresses a spatial relationship, e.g., "in the north part of", "near", "adjacent to". It can be categorized into monadic, dyadic operators and so on. The priority of $\oint$ is only below $\neg$. It should be noted that the number of a spatial operator's operands is one less than that of the corresponding spatial predicate. For example, the phrase "adjacent to" is a dyadic spatial relationship predicate, but as a spatial relationship operator it is monadic. Theoretically, according to the logic inference rules it is allowed to nest spatial relationship operator $\oint$ with other geo-spatial propositions, although it is quite rare in natural languages.

The inference rules are defined as follows:
(1) A single place name, or a single place name with a geo-spatial predicate, or a single name's negation is an atomic geo-spatial proposition;
(2) If $\oint$ is a monadic spatial relationship operator and $p$ is a place name, $\oint p$ is a geo-spatial proposition sentence;
(3) If $\oint$ is a dyadic spatial relationship operator and both $p$ and $q$ are place names, $p \oint q$ is a geo-spatial proposition sentence. Multivariate relationship operators can be inferred in the same way;
(4) If $P$ is a geo-spatial proposition sentence, its negation $\neg P$ is a geo-spatial proposition sentence;
(5) If $P$ and $Q$ are two geo-spatial proposition sentences, their conjunction $P \wedge Q$ is a geo-spatial proposition sentence;
(6) If $P$ and $Q$ are two geo-spatial proposition sentences, their disjunction $P \vee Q$ is a geo-spatial proposition sentence.

The spatial relationships can be classified into topological relationship, metric relationship and directional relationship. Cognitive spatial relations are predominantly topological, but metric factors such as distance and direction often refine the relations and characterize prototypical relations [18]. Topological spatial relationships can be characterized by 9-intersection model or RCC model. Both metric and directional relationships can be characterized by qualitative or quantitative models. In previous research, the representation models and inference rules of spatial relationship predicates are defined [19], so the set of spatial relationship operators can build up based on them. Although spatial relationship operators support quantitative representation of metric and directional relationships, qualitative representation is still the main method in natural language and web documents.

Similar to thematic information, geographic information in information units of a document is the disjunctions of atomic geo-spatial propositions, and geographic information in queries is the combination of atomic geo-spatial propositions by Boolean logic.

Qualitative representation of geo-spatial information relies on geographic knowledge base (GKB). GKB maintains knowledge of geographic entities in the real world. It includes geographic entities, their types, properties and relationships. There are three types of GKB in GIR: gazetteer, geographic thesauri and geographic ontology. It can help us to identify place names, disambiguate place names and confirm the geographic extent of place names [20]. For the qualitative geographic representation model, place names in the GKB formulate all possible place names in the geo-spatial propositions. The spatial relationships and type relationships between geographic entities, which are stored explicitly in the GBK, are the basis for geospatial proposition inference.

## 4. Similarity measurement based on qualitative representation model

### 4.1 Similarity measurement model

In a GIR system, the similarity between documents and queries is measured to rank candidate documents. Both thematic and spatial similarity should be taken into account synthetically. Based on the qualitative representation model described above, a semantic similarity measurement method is proposed based on evidence theory and fuzzy logic.

For a user, the importance of a document is how much it can satisfy the query. Note given a document $d$ and a query $q$, that the degree that $q$ satisfies $d$ is not equal to the degree that $d$ satisfies $q$. The relevance between a document and a query is direction dependent, as $d \rightarrow q$. As a result, the process of similarity reasoning is also directional, which is the reasoning belief of $d \rightarrow q$. For instance, when a query is "western restaurant", the result "pizza in New York" is reasonable. On the contrary, when a query is "pizza in New York", the result "western restaurant" is not quite satisfied. Hence the reasoning direction should be taken into consideration to evaluate the similarity.

On the basis of qualitative representation model, a document $d$ is composed of $n$ information units. Then the similarity measurement can be divided into two steps:

(1) Calculate the similarity between every information unit $u_i$ and a sub query $q$, denoted as $Ru(u_i,q)$. It is the combination of thematic and geographic similarity between $u_i$ and $q$;
(2) Combine the similarities of all information units in the document $d$ with the query $q$ and get the relevance of document $d$ with $q$, denoted as $Rq(d, q)$ where $k$ is the combination function:

$$Rq(d,q_i) = k(Ru(u_1,q), Ru(u_2,q), \ldots, Ru(u_n,q)) \qquad (4)$$

In this method, two kinds of similarities are combined in different ways. A document is a set of information units taken as evidences, so that their similarities are combined by evidence theory. On the other side, the thematic and spatial information of information units, and the query are all represented by proposition logic. We thus introduce fuzzy logic reasoning for the similarity combination function.

### 4.2 Similarity between an information unit and a query

Since a query can be represented as a simple information unit, the similarity $Ru(u_i, q)$ between an information unit $u_i$ and a query $q$ can be converted into the similarity between two information units. Each information unit is composed of thematic and spatial information as defined above, so the similarity measurement can be divided into three parts: thematic similarity, spatial similarity and their combination.

In the qualitative representation model, all the elements, including thematic information, spatial information and queries are represented by propositional logic in which propositional variants are connected by logic operators {∧, ∨, ¬}. Hence, fuzzy logic reasoning is employed to combine the similarities between propositional variants. Fuzzy logic [21] deals with reasoning that is approximate rather than fixed and exact. The reasoning is based on rules. The precondition of a rule is the combinations of fuzzy sets built up by logic operators, and the conclusion is the fuzzy set with its corresponding membership function.

Fuzzification, which assigns a fuzzy set to each input variant, should be conducted before rules matching. Each input variant is associated with one or multi linguistic variables (presumes or propositions), and its membership is given. If there are multi variants when matching rules, the conjunction and disjunction of these variants take the minimum operator and the maximum operator respectively [22]. Assume that proposition variants $A$ and $B$ are two fuzzy sets on domain $X$. Their membership functions are $\mu_A(x)$ and $\mu_B(x)$, $x \in X$. There will be the following conclusions.

The negation of $A$ is the complementary set of its fuzzy set:
$$\mu_{\bar{A}}(x) = 1 - \mu_A(x) \tag{5}$$

The conjunction of $A$ and $B$ is the intersection of their fuzzy sets, and the result membership is the minor of that of $A$ and $B$:
$$\mu_{A \cap B}(x) = \min(\mu_A(x), \mu_B(x)) \tag{6}$$

The disjunction of $A$ and $B$ is the union of their fuzzy sets, and the result membership is the larger of that of $A$ and $B$:
$$\mu_{A \cup B}(x) = \max(\mu_A(x), \mu_B(x)) \tag{7}$$

Each matched rule reveals its support to the conclusion. Finally all matched rules are combined together to generate a definite output value. The combination function can be weighted average method or others.

**4.2.1 Thematic similarity**

The thematic similarity between two information units depends on whether and how much they convey the same thematic information. The measure is a function of thematic keywords in the two information units. It is a measurement of the relevance in semantics rather than the co-occurrence frequency of same keywords, which is used in conventional information retrieval models. Moreover, the frequency of keywords in information units may be small in the qualitative representation model, so it is not so useful for the measurement of thematic similarity.

According to the qualitative representation model, thematic information is a proposition of thematic keywords. Assume $a$ and $b$ are two thematic keywords, and their thematic relevance operator is denoted as $\oplus$, then:
$$\begin{cases} a \oplus b = h(a, b) \\ a \oplus \neg b = 1 - h(a, b) \end{cases} \tag{8}$$

where $h(a,b)$ is the conceptual similarity function. Since the similarity is direction dependent, the operator $\oplus$ does not satisfy the commutative law, which is $a \oplus b \neq b \oplus a$.

In an information unit, the thematic information $X$ is composed of thematic keywords {$x_1$, $x_2$, …, $x_n$} that are connected by disjunctions. If $a$ is a keyword, then the thematic similarity between $X$ and $a$ is:

$$\begin{cases} X \oplus a = max\{x_i \oplus a | i = 1,2,...,n\} \\ X \oplus \neg a = 1 - min\{x_i \oplus a | i = 1,2,...,n\} \end{cases} \quad (9)$$

According to the max-min method, the thematic similarity between the thematic information $X=\{x_1, x_2, ..., x_n\}$ and $Y=\{y_1, y_2, ..., y_k\}$ is:

$$X \oplus Y = min\{X \oplus y_i | i = 1,2,...,k\} \quad (10)$$

If $H=\{h_1, h_2\}$ is the thematic information of a query with two keywords, and ⊌ is the logic operator connecting the sub propositions that can be ∨ or ∧, then the thematic similarity between the information unit and the query can be measured as:

$$X \oplus H = \begin{cases} max\{X \oplus h_1, X \oplus h_2\}, & ⊌ = \vee \\ min\{X \oplus h_1, X \oplus h_2\}, & ⊌ = \wedge \end{cases} \quad (11)$$

Any complicated proposition can always be decomposed into the forms above, so that the semantic similarity measurement $h(a,b)$ between two keywords is the essential problem.

Relationships between concepts in domain ontology provide a solution to measure the semantic similarity between two keywords. So we propose a method based on the weighted minimum distance of concepts [23] to calculate the semantic similarity between thematic keywords. Concepts in a domain ontology and their connections form a tree or a net structure, where concepts are nodes and connections are links. And BT (broader term) relationship, NT (narrower term) relationship and RT (related term) relationship are three important relationships in them. The shortest path between two concepts in a conceptual sematic net is the path that goes through least nodes from start node to end node. According to the relevance between concept nodes, each link is set a weight. The weighted shortest path is the shortest path that has the minimum sum of link weights through the path.

If $a$ and $b$ are two thematic keywords corresponding to two concepts in ontology, their weighted shortest path $TD(a,b)$ is defined as follows:

$$TD(a,b) = \left(\frac{C_{a,x_1}}{L_{x_1}} + \frac{C_{x_1,x_2}}{L_{x_2}} + \cdots + \frac{C_{x_{n-1},x_n}}{L_{x_n}} + \frac{C_{x_n,b}}{L_b}\right) \quad (12)$$

$C_{i,j}$ is the weight linking adjacent concepts $i$ and $j$ and is assigned by the relevance from $i \to j$. $L_i$ is the depth of concept $i$ in the concept ontology tree. The expression $a \to x_1 \to x_2 \to ... \to b$ is the shortest path from concept $a$ to concept $b$. The final semantic similarity is got by normalizing the formula result above. The normalization can use the longest distance of two concepts denoted as $MD$, or logarithm methods. So the conceptual similarity between two thematic keywords $a$ and $b$, i.e. $h(a,b)$ can be:

$$a \oplus b = h(a,b) = 1 - \frac{TD(a,b)}{MD} \quad (13)$$

or

$$a \oplus b = h(a,b) = \frac{1}{1+\ln(1+TD(a,b))} \quad (14)$$

In practice, different weights need to be assigned to NT, BT, and RT relationships. Tudhope and Taylor [23] suggested that the distance of two concepts connected by RT relationship are farther than that of NT or BT relationships, which can be represented as $w_{NT}=w_{BT}\leq w_{RT}$. Their exact values are experimental, for example, $w_{NT}=w_{BT}=0.5$ and $w_{RT}=0.8$. The values are not that important since they are just references for ranking by the relevance of semantics. But Tudhope's method does not take the similarity direction into consideration. The similarity is direction dependent and the direction presents whether and how much the information, which one concept conveys can overlap another. Semantic similarities of *a→b* and *b→a* are different. So we assign BT and NT relationships different weights, where NT weight is minimum, BT weight is larger, and RT weight is the largest. Hence we have $w_{NT} < w_{BT} < w_{RT}$.

**4.2.2 Spatial similarity**

The spatial similarity between two information units is the semantic relevance of their spatial information. The spatial information in information units is a sentence of atomic geo-spatial propositions, so the similarity is also a function of them.

Suppose *p* and *o* are two atomic geo-spatial propositions, and their spatial relevance operator is $\odot$, then according to fuzzy logic, there will be:

$$\begin{cases} p \odot o = g(p, o) \\ p \odot \neg o = 1 - g(p, o) \end{cases} \quad (15)$$

where *g(p,o)* is the spatial similarity function. The spatial similarity is also directional as well as the thematic similarity, so the operator $\odot$ does not satisfy commutative law, which is $p \odot o \neq o \odot p$.

A compound geo-spatial proposition *c* can be seen as the combination of two geo-spatial propositions, namely $c = c_1 \uplus c_2$, where $\uplus$ is $\vee$ or $\wedge$. Then the spatial similarity between proposition *p* and compound proposition *c* is as the following formula, and any complicated compound propositions can break down into the form.

$$p \odot c = \begin{cases} max\{p \oplus c_1, p \oplus c_2\}, & \uplus = \vee \\ min\{p \oplus c_1, p \oplus c_2\}, & \uplus = \wedge \end{cases} \quad (16)$$

Then, the spatial information *S* in an information unit of a document is composed of a geo-spatial proposition set {$s_1$, $s_2$, …, $s_n$} in which elements are connected by disjunctions. Assume *p* is a geo-spatial proposition, the spatial similarity between *S* and *p* will be:

$$\begin{cases} S \odot p = max\{s_i \odot a | i = 1, 2, …, n\} \\ S \odot \neg p = 1 - min\{s_i \odot a | i = 1, 2, …, n\} \end{cases} \quad (17)$$

Furthermore, based on max-min method, the spatial similarity between two pieces of spatial information $S=\{s_1, s_2, …, s_n\}$ and $T=\{t_1, t_2, …, t_k\}$ is:

$$S \odot T = min\{S \odot t_i | i = 1, 2, …, k\} \quad (18)$$

Suppose *G* is a compound geo-spatial query proposition, which is composed of two sub propositions {$g_1$, $g_2$} connected by logic operator $\uplus$ ($\vee$ or $\wedge$). Then the spatial similarity between an information unit in a document and a query would be:

$$S \odot G = \begin{cases} max\{S \oplus g_1, S \oplus g_2\}, & \uplus = \vee \\ min\{S \oplus g_1, S \oplus g_2\}, & \uplus = \wedge \end{cases} \quad (19)$$

Then the function *g(p,o)* is the key problem. It can be determined by the belief between atomic geo-spatial propositions, which is inferred based on geographic knowledge base and qualitative spatial reasoning. A Geographic knowledge base generally stores the name, geographic extent, and entity type of geographic entities. And it stores spatial relationships explicitly, especially the whole-part relationship. Geographic knowledge base is the fundamental of spatial reasoning and computing.

Generally, geo-spatial propositions in a document mostly are simple place names, some of which are constrained by spatial predicates. The same is true for geographic information queries. Hence, the similarity measurement of geographic information turns to the semantic similarity between two place names.

The semantics of place names includes conceptual and locational features, so the semantic similarity between place names should take both features into account. For the conceptual similarity, we use Jones' algorithm [24] that calculates the distance of concepts' depths in an ontology tree based on geographic knowledge base. Then the conceptual similarity is determined by:

$$CS(p,o) = 1 - \left(\sum_{x \in \{p.PartOf - o.PartOf\}} \frac{\alpha}{L_x} + \sum_{y \in \{o.PartOf - p.PartOf\}} \frac{\beta}{L_y} + \sum_{z \in \{p,o\}} \frac{\gamma}{L_z}\right) \quad (20)$$

$CS(p,o)$ is the conceptual similarity between the place names $p$ and $o$. $L_x$, $L_y$ and $L_z$ are the depths of the sets of places $x$, $y$, $z$ in the ontology tree respectively. The sets of terms $p.partOf$ and $o.PartOf$ refer to the transitive closure of the parents of $p$ and $o$ respectively in the ontology. The weights $\alpha$, $\beta$ and $\gamma$ are harmonic coefficients, providing control over the application of the measure. Generally, set $\alpha=\beta=1.0$, and when $p$ and $o$ are siblings in the tree, set $\gamma=1.0$, otherwise set $\gamma=0$.

The locational similarity is measured based on the rule "topology matters, metric refines" [18]. When the spatial extents of place name $p$ and $o$ are large (province or larger than a province), the measurement can take topological semantic relevance [25] directly and does not need refinement by distance. When the spatial extents are small (city or smaller than a city), the measurement can take Andrade and Silva's algorithm [26], which takes both topology and metric into consideration as follows:

(1) Topology similarity

$$Inclusion(p,o) = \begin{cases} \frac{NumDescendant(o)+1}{NumDescendants(p)+1} & o \subseteq p \\ 0 & others \end{cases} \quad (21)$$

(2) Metric similarity

$$Proximity(p,o) = \frac{1}{1+Distance(p,o)/diagnoal(p)} \quad (22)$$

(3) Judge if place names are siblings

Set $Sibling(p,o)=1$, when their parent set are same, otherwise, set $Sibling(p,o)=0$.

(4) Combine three values above to generate the locational similarity $LS(p,o)$

$$LS(p,o)=b \times \{Inside(p, o)+Proximity(p,o)\}+(1-b) \times Siblings(p,o) \quad (23)$$

where $b$ is a harmonic coefficient ranging from 0 to 1.

The two similarities, $LS$ and $CS$, can be combined into a weighted function as the spatial similarity $g(p,o)$ as follows:

$$g(p,o)=w_g \times LS(p,o)+w_h \times CS(p,o) \quad (24)$$

where $w_g$ and $w_h$ are harmonic coefficients of the $LS$ and $CS$ respectively, ranging from 0 to 1. Jones [24] et al set $w_g$ to 0.6 and $w_h$ to 0.4. In practice, all the harmonic coefficients above are relevant to the spatial extents of queries and documents.

When geo-spatial propositions are too complicated, the qualitative reasoning method cannot assess reasoning belief between arbitrary geospatial propositions. The propositions should thus be converted into geo-spatial extent with coordinates, and the spatial similarity is calculated based on quantitative methods, such as overlapping area or Euclidean distance.

### 4.2.3 Thematic and spatial similarities combination

At last, we combine thematic similarity and spatial similarity together to get the final similarity. Similarity $Ru(u_i, q)$ between an information unit $u_j$ in a document and a query $q$ is as follows:

$$Ru(u_i, q)=h(Ru_g(u_i, q), Ru_t(u_i, q)) \quad (25)$$

where $Ru_g$ is spatial similarity, $Ru_t$ is thematic similarity, and $h$ is a combination function that can be geometric average, arithmetic average, mathematic product, weighted arithmetic average, Euclidean distance. Yu and Cai [27] proposed a dynamic document ranking scheme to combine the two similarities on a per-query basis. The authors introduced query specificity, including both thematic specificity and geographic specificity, to determine the relative weights of different sources of ranking evidence for each query. It is a better but more complex solution. As in this paper, geometric average is a good choice if ignoring query specificity, so it is used in our experiment too.

### 4.3 Similarity between a document and a query

The measurement of similarity $Rq(d, q)$ between the document $d$ and the query $q$ is modeled as an uncertainty reasoning process combining the similarities between all information units in $d$ and $q$. Each information unit can be a piece of evidence that supports the relevance between a document and a query, so that the similarity can be evaluated as the belief that how much the evidence supports the proposition "document $d$ is relative to query $q$". The similarity measurement between a document and a query meets the preconditions of evidence theory. First, the similarity between an information unit and a query is not an exact estimation of a random event. It satisfies the coincidence proportion of information or the reasoning belief more than the Bayesian probability. Second, each information unit is independent from others, and the order that information units occurs will not influence the similarity measures. So the evidence theory can be used to measure the similarity between a document and a query. We take D-S evidence theory to combine evidence belief and use the combined belief to judge the relevance between $d$ and $q$. Dempster-Shafer evidence theory (D-S theory for short) is the mathematical theory about evidence [28]. With given evidence, D-S theory defines the belief measures for every proposition set (assumptions to be proved). It supports one piece of evidence to assess beliefs for multi propositions, and can combine beliefs of much evidence.

In the D-S theory, a complete set composed of atomic propositions that are mutually exclusive is called identical framework, denoted as $\Theta$, which represents all possible answers for some problem but only one is right. The power set of $\Theta$, denoted as $2^\Theta$, is the set of all the sub set where each sub set is a proposition. There are three important functions in D-S theory. For a given proposition (or a set) $A$ in $2^\Theta$, the basic probability assignment function $m(A)$ measures the proportion of all relevant and available evidence supporting the proposition $A$. The belief function $Bel(A)$ expresses the belief of the proposition $A$. And the probability function $Pl(A)$ expresses the plausibility that proposition $A$ is not false. $Bel(A)$ and $Pl(A)$ are the lower and upper bounds respectively of the interval containing the precise probability of the proposition $A$. These functions meet:

$$Bel(A) = \sum_{B|B\subseteq A} m(B) \tag{26}$$

$$Pl(A) = \sum_{B|B\cap A \neq \phi} m(B) \tag{27}$$

where the function $m$ meets:

$$\begin{aligned} m: 2^\Theta &\to [0,1] \\ 0 \leq m(x) &\leq 1, \forall x \in 2^\Theta \\ \sum_{x\in 2^\Theta} m(x) &= 1 \\ m(\phi) &= 0 \end{aligned} \tag{28}$$

When combining a number of evidences, D-S theory provides a combination function as follows:

$$m_{i+1}(C) = \frac{\sum_{A\cap B=C} m_i(A) m^{i+1}(B)}{\sum_{A\cap B\neq \phi} m_i(A) m^{i+1}(B)} \tag{29}$$

where $m_i$ is the value of $m$ before the $(i+1)$th evidence occurs, $m^{i+1}$ is the belief assignment of the $(i+1)$th evidence for every proposition, and $m_{i+1}$ is the value of function $m$ after the $(i+1)$th evidence is combined.

Following Equations 26 and 27, equation 29 yields an probability interval [$Bel(A)$, $Pl(A)$] for every possible proposition combinations, which is the range for the probability $p(A)$ where $A$ is true based on all the given evidence. Formally, when D-S theory is applied in qualitative reasoning, two propositions need to be proved: *T=document is relative to query*, *F=document is not relative to query*. Then the identification framework are $\Theta=\{T, F\}$, and its power set is:

$$2^{\Theta} = \{\phi, \{T\}, \{F\}, \{T,F\}\} \tag{30}$$

where $m(\phi) = 0$. Propositions $\{T, F\}$ allow uncertain estimation in the proving process.

Assume that the similarity between information unit $u_j$ and query $q$ is $\alpha$ that is got by the method described in section 4.2. If the information unit is seen as evidence, then the belief it supports proposition $\{T\}$ is $\alpha$. There is no evidence to prove proposition set $\{F\}$ directly. According to the combination theory, the best way is to assign probability $1-\alpha$ to the whole set $\{T, F\}$. Then we get:

$$Ru(u_i, q) = \alpha \Rightarrow \begin{cases} m(\{T\}) = \alpha \\ m(\{F\}) = 0 \\ m(\{T,F\}) = 1 - \alpha \\ m(\{\phi\}) = 0 \end{cases} \tag{31}$$

Then $Rq(d, q)$ is solved iteratively based on $Ru(u_i, q)$. All information units in a document are added as evidence in order, and the basic probability assignment of every piece of evidence is calculated. Then the $m$ function of every proposition set is updated after combining the belief of this evidence by D-S theory (Table 1).

Table 1 Computation of the evidence combination algorithm

|  | $m^{i+1}(\{T\})$ | $m^{i+1}(\{T,F\})$ |
|---|---|---|
| $m_i(\{T\})$ | $m_i(\{T\}) \cdot m^{i+1}(\{T\})$ | $m_i(\{T\}) \cdot m^{i+1}(\{T,F\})$ |
| $m_i(\{T,F\})$ | $m_i(\{T,F\}) \cdot m^{i+1}(\{T\})$ | $m_i(\{T,F\}) \cdot m^{i+1}(\{T,F\})$ |

Here $m_i$ is the value of $m$ function after combining $i$ pieces of evidence. $m^{i+1}$ is the basic probability assignment after $(i+1)$th piece of evidence added. If the intersection set is null, its results should be removed and the value of $m$ should be normalized again.

After all evidence is added, the $m$, $Bel$ and $Pl$ function values of every proposition set are obtained. The probability $p(T)$ that supports the proposition $T$ and the probability $p(F)$ that supports the proposition $F$ meet the formulas below:

$$Bel(\{T\}) \leq p(T) \leq Pl(\{T\}), \ Bel(\{F\}) \leq p(F) \leq Pl(\{F\}) \tag{32}$$

$Bel(\{T\})$, $Pl(\{T\})$, $Bel(\{F\})$ and $Pl(\{F\})$ are basic indicators for the similarity between a document and a sub query. The similarity can be defined as $Bel(\{T\})$ directly or as follows:

$$Rq(d, q_j) = \frac{Bel(\{T\})}{Bel(\{F\})}, \text{ or } Rq(d, q_j) = 1 - \frac{1}{\ln\left(e + \frac{Bel(\{T\})}{Bel(\{F\})}\right)} \tag{33}$$

In conclusion for this section, our similarity metrics clearly separate the process into two steps. Firstly, it calculates the similarity between an information unit and a query by fuzzy logic. Then it calculates the similarity between a document, which is composed of information units, and a query based on evidence theory. The processes define clear mathematic method and can get results correctly. To testify our method, an experiment is conducted in the next section.

## 5. Experiment and Discussion

## 5.1 Experiment

An experiment is conducted to verify the qualitative representation and similarity measurement method. All candidate documents are collected from the homepage of Chinese Mining Industry (http://www.chinamining.com.cn), which is written in Chinese natural language. The contents are all about geology and mineral resources. In such a restricted domain, limited and effective domain ontology can be built easily. The ontology is used as the thematic knowledge base in the search process. The extraction of information units is taken with the help of manual work. A document is segmented into several independent paragraphs by manual work, and then a Chinese word segmentation tool is used to handle each paragraph to extract thematic information and geographic information. Through the work above, 100 documents are collected as the test set. Each document has 2 to 3 information units in average, and every information unit has a thematic keyword and a place name representing the geographic location.

For instance, the statement "Chromite distributes in 13 provinces in China, mainly in Tibet, Inner Mongolia, Xinjiang and Gansu. Bauxite distributes in 20 provinces, mainly in Shanxi, Guizhou, Henan and Guangxi" can be represented as two information units, namely:

$d$ = { ({Chromite}, {Tibet, Inner Mongolia, Xinjiang, Gansu}),
({Bauxite}, {Shanxi, Guizhou, Henan, Guangxi}) }

A simple mineral ontology is built for the documents (Fig. 1). It includes 1 primary mineral type, 3 secondary level mineral types, 73 third level mineral types, 47 fourth level mineral types and 17 fifth level mineral types.

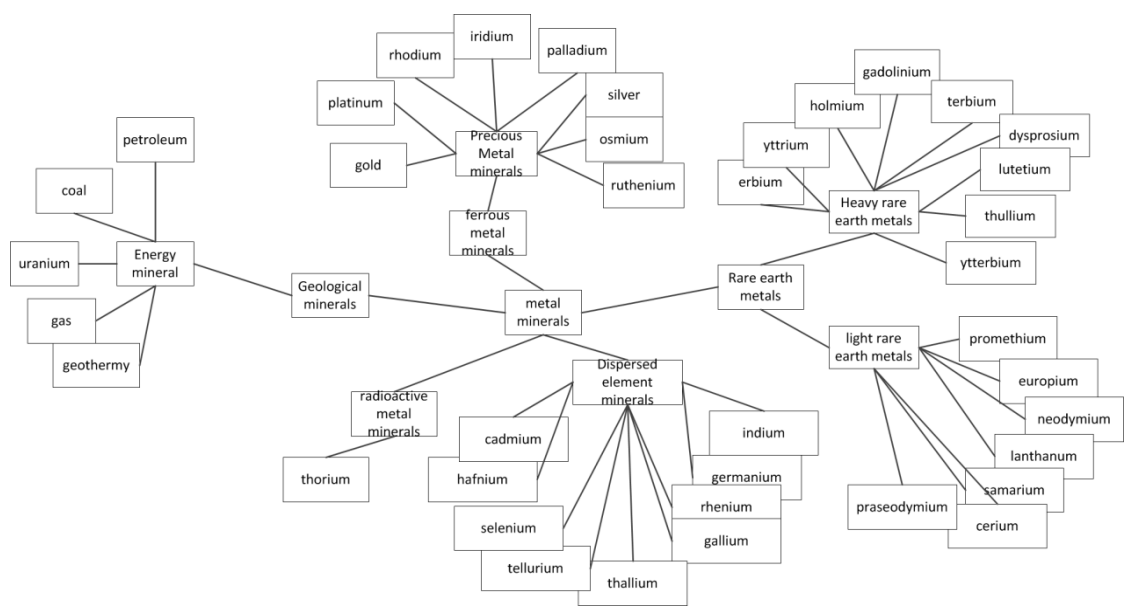

Fig. 1 Mineral ontology used in the experiment

A gazetteer covering whole Chinese county regions is used in the experiment. It includes the place names of 34 provinces, 360 cities, 2939 counties and 20266 towns. Conventional GIR method is adopted to compare with the qualitative method presented in this paper. It uses the vector model to calculate the thematic similarity, and uses the overlapping area proposition of bounding boxes as the geographic similarity, and then combines them by the geometric average.

Ten queries are conducted on the collected documents by the qualitative method and the conventional method. They are quite usual forms that people often use, like "precious metals in Hebei Province", "metals in Shijiazhuang City" etc. The relevant document set is firstly manually chosen by several people and then picked up from these candidates. The results are assessed by 11 points standard recall rates in which we take 11 standard recall levels (0.0, 0.1,…, 1.0) denoted as $r_j$ and make $P(r_j)=max(P(r))$ $(r_j<=r)$. Figure.2 presents the mean precision of the ten queries and it shows that the qualitative method can get better searching precision and can detect small differences between documents. Additionally, there are less mismatch documents than the conventional method.

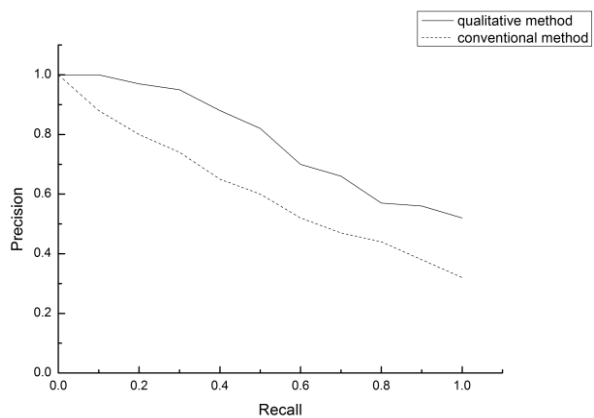

Fig.2 Comparison of 11-points recall rates of two GIR methods

## 5.2 Discussion

The qualitative method can get a better result than conventional GIR in the situations above. The proposition representation and corresponding reasoning rules fit these large space scale queries better. For small space scale query like a place inside a city or town, conventional GIR can give people more exact results and performs better. It is a good idea to combine the two kinds of method together in a system and use them in different situations. When the query's space scale is large (larger than a city or province), it is better to use the qualitative method since it is more adaptive to people spatial cognition and can get results more easily and quickly. When the query's space scale is moderate (like a county), we take qualitative and quantitative methods in different levels. The qualitative method is used firstly to filter documents fitting the space extents, and then the quantitative methods are used to handle the filtered documents more exactly. When the query's space scale is small (like a street or block), it is quite better to use the quantitative methods to get a more exact result.

## 6. Conclusion

This paper proposes a new GIR representation method and the corresponding measurement of similarity based on proposition logic and fuzzy reasoning method, providing a new way to improve the efficiency of GIR. Compared with conventional GIR, the qualitative method takes into account semantics and uses geographical ontology or gazetteer directly so that spatial information needs not to be simplified. It can thus represent documents' information and users' queries objectively. The matching method based on real information contents makes the retrieved documents more satisfactory to common knowledge organization than conventional quantitative method especially in the large spatial scale. Besides, the qualitative model also supports the quantitative methods. In the future, we will study further on the information units' extraction, improve qualitative reasoning assessing model and build a more complete geographic knowledge base.

**Acknowledgments**

This work is supported by The National Natural Science Foundation of China (41271385, 41171296).